# Dynamic Bowtie for Fan-beam CT

Fenglin Liu[1,2], Ge Wang[2]*, Wenxiang Cong[2], Scott Hsieh[3], Norbert Pelc[3]*

[1] Engineering Research Center of Industrial Computed Tomography Nondestructive Testing, Key Lab of Optoelectronic Technology and System, Ministry of Education, Chongqing University, Chongqing 400044, China

[2] School of Biomedical Engineering and Sciences, Virginia Tech, Blacksburg, Virginia 24061, USA

[3] Department of Biomedical Engineering, Stanford University, Stanford, CA 94305, USA

**Abstract.** A bowtie is a filter used to shape an x-ray beam and equalize its flux reaching different detector channels. For development of spectral CT with energy-discriminative photon-counting (EDPC) detectors, here we propose and evaluate a dynamic bowtie for performance optimization based on a patient model or a scout scan. Our dynamic bowtie modifies an x-ray beam intensity profile by mechanical rotation and adaptive adjustment of the x-ray source flux. First, a mathematical model for dynamic bowtie filtering is established for an elliptical section in fan-beam geometry, and the contour of the optimal bowtie is derived. Then, numerical simulation is performed to compare the performance of the dynamic bowtie in the cases of an ideal phantom and a realistic cross-section relative to the counterparts without any bowtie and with a fixed bowtie respectively. Our dynamic bowtie can equalize the expected numbers of photons in the case of an ideal phantom. In practical cases, our dynamic bowtie can effectively reduce the dynamic range of detected signals inside the field of view. Although our design is optimized for an elliptical phantom, the resultant dynamic bowtie can be applied to a real fan-beam scan if the underlying cross-section can be approximated as an ellipse. Furthermore, our design methodology can be applied to specify an optimized dynamic bowtie for any cross-section of a patient, preferably using rapid prototyping technology. This fan-beam dynamic bowtie work could be extended to the cone-beam geometry in a follow-up study.

Keywords: Dynamic bowtie, tube current modulation, elliptical phantom, computed tomography (CT)

## 1. Introduction

A modern x-ray computed tomography (CT) scanner always uses a beam shaping component, which is referred to as a bowtie filter for its bowtie-like shape. A bowtie filter is normally a piece of material, which is placed between an x-ray source and an object to be imaged. Such a beam shaping device also helps to improve image quality. Since the intensity readings of the x-rays measured by the detectors are greatly equalized, the dynamic range is reduced for a data acquisition system to extract inherent information effectively [1,2]. By blocking low energy x-rays, the bowtie also works with an x-ray beam filter to reduce the so-called beam-hardening effect [3]. By blocking radiation to the periphery of the patient where attenuation is lowest, the scatter-to-primary ratio is reduced. Generally, a bowtie reduces the radiation dose to a patient, especially in the peripheral region of a field-of view (FOV) [4-6].

In recent years, there is a gradually increased interest in precisely manipulated tomographic settings, such as a fine adjustment of the x-ray intensity distribution upon an x-ray detector array. Ideally, this adjustment should be individualized to maximize the information content in the projection domain. Of particular interest, recently developed energy-discriminative photon-counting (EDPC) detectors can resolve photons and their associated energies [7-9]. This photon-counting detector technology enables the development of the next generation CT systems for spectral or true-color imaging, being instrumental in functional, cellular and molecular studies. However, the dynamic range of the EDPC detector is rather limited, and prefers relatively low photon flux rates. This requirement underlines the importance of a dynamic bowtie design for the dynamic range of the EDPC detectors to cover that of projection data, defining conformal CT protocols.

The primary function of a dynamic bowtie is to smartly filter the radiation emitted towards an object in synchrony with a high-speed data acquisition process. It is desirable to have the dynamics of the bowtie guided by a patient model or a scout scan. Various beam-shaping filters were reported to shape the x-ray beam dynamically [10-12]. The effects of the bowtie shape on image quality and radiation dose were also investigated [13-17]. However, these dynamic bowtie designs are not sufficiently mature, and currently commercial CT scanners use only fixed bowtie. The challenges for a dynamic bowtie design include selection of an appropriate bowtie shape to match projections of a patient and synchronization of the bowtie with the other system components for data acquisition.

In this paper, we propose a novel dynamic bowtie design, which modifies an x-ray beam intensity profile by mechanical rotation of the bowtie in a conformal shape and, at the same time modulates the x-ray tube current [18-20]. In this initial investigation, the dynamic bowtie targets the fan-beam geometry, and will be extended to the cone-beam geometry in a follow-up project. In the next section, the contour profile of the bowtie is derived assuming a controllable tube current. In the third section, numerical data are described to demonstrate the performance of the dynamic bowtie. In the last section, relevant issues are discussed.

## 2. Methods

When an x-ray fan-beam irradiates an object, the path length of each x-ray varies significantly within the fan-beam and as a function of view angle. The large variations result in data overflow, which is especially problematic for spectral detectors. To address this issue, a smart bowtie can be used to optimally shape the x-ray beam so that the expected numbers of photons are equalized across detector channels and view angles. The major task of the dynamic bowtie design is to determine the bowtie shape and modulate the source current for dynamic adaptation to the path length variations during a CT scan.

### 2.1. Scanning Geometry

Since a cross-section of head, chest, and abdomen can be approximated to be elliptical, the bowtie filter can be analytically designed for such objects, which can be defined with semi-major axis $A$ and semi-minor axis $B$ as follows:

$$f(x,y) = \begin{cases} \rho & \text{for } \frac{x^2}{A^2} + \frac{y^2}{B^2} \leq 1 \\ 0 & \text{otherwise} \end{cases} \tag{1}$$

We assume the fan-beam geometry with detector cells distributed at equiangular intervals, as shown in Figure 1. An x-ray source $S$ is rotated along a circular trajectory of radius $R_0$, and $\varphi$ is the polar angle relative to the x-axis to represent the projection view of the fan beam. Let $L(\varphi,\gamma)$ denotes a ray in the fan-beam, where the angle $\gamma$ specifies a ray within the fan beam. Then, the projection $P(\varphi, \gamma)$ of x-ray fan-beam through the homogenous elliptical object with an attenuation coefficient $\mu$ can be formularized as [21]

$$P(\varphi,\gamma) = \begin{cases} \frac{2\mu AB}{a^2(\varphi+\lambda)}\sqrt{a^2(\varphi+\gamma) - (R_0 \sin(\gamma))^2}, & \text{for } |R_0 \sin(\gamma)| \leq a(\varphi+\gamma) \\ 0, & |R_0 \sin(\gamma)| > a(\varphi+\gamma) \end{cases}, \tag{2}$$

where $a^2(\varphi+\gamma) = A^2\cos^2(\varphi+\gamma) + B^2\sin^2(\varphi+\gamma)$.

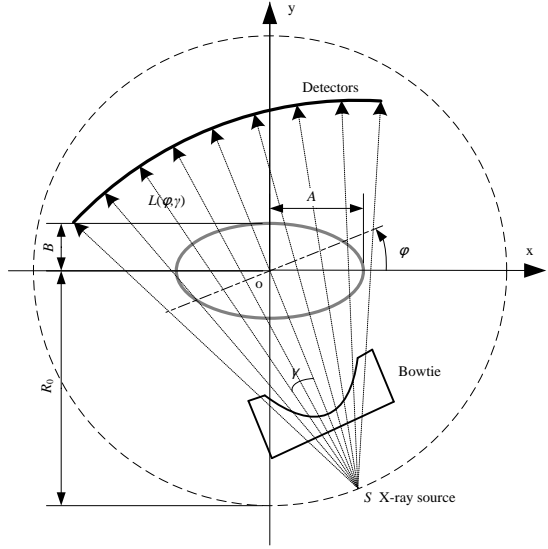

Figure 1: Equiangular fan-beam geometry.

## 2.2. Bowtie Design

The new bowtie is rotated around the axis orthogonally through its center to compensate for the variation of projection data $P(\varphi, \gamma)$ as a function of the fan-beam view angle $\varphi$ for the minimization of the dynamic range of detected signals. However, the central thickness of the rotating bowtie is constant. Hence, according to the view-angle-dependent variation of the patient thickness, we can accordingly modulate the current of the x-ray source [18,20] to keep the expected number of photons detected by the central detector the same during the CT scan.

Let a function $N_i(\varphi)$ denotes the expected number of photons emitted from the x-ray source at the scanning angle $\varphi$. Let $N_o(\varphi, \gamma)$ be the expected number of photons which pass through an object and hit the detectors. By Beer-Lambert's law, we have

$$N_o(\varphi, \gamma) = N_i(\varphi) e^{-B(\varphi,\gamma)} e^{-P(\varphi,\gamma)}, \tag{3}$$

where $B(\varphi, \gamma)$ is the absorbance function determined by the bowtie. Then, the number of photons at the central detector can be written as

$$N_o(\varphi, 0) = N_i(\varphi) e^{-(B(\varphi,0)+P(\varphi,0))}. \tag{4}$$

To keep the central thickness of the bowtie and the number of photons at the central detector unchanged for different projection view angles, we can modulate the current of the x-ray source $N_i(\varphi)$ so that

$$B(\varphi, 0) = \ln\left(N_i(\varphi) e^{-P(\varphi,0)}\right) + \text{constant}. \tag{5}$$

Then, we have

$$N_o(\varphi,\gamma) = N_o(\varphi,0)e^{(B(\varphi,0)+P(\varphi,0)-B(\varphi,\gamma)-P(\varphi,\gamma))} \quad . \tag{6}$$

The proposed bowtie function is to minimize the dynamic variation of detected signals for all projection views. Ideally, $N_o(\varphi,\gamma)$ and $N_o(\varphi,0)$ should be constant for all $\varphi$ and $\gamma$. Hence, we have the following bowtie function:

$$B(\varphi,\gamma) = P(\varphi,0) - P(\varphi,\gamma) + B(\varphi,0) + \text{constant}. \tag{7}$$

Clearly, the bowtie function can be determined from the involved projections, according to Eq. (7). Since $B(\varphi, \gamma)$ changes with $\varphi$ periodically, it is feasible to link $B(\varphi, \gamma)$ to $P(\varphi, \gamma)$ in synchrony with the source rotation. In other words, the 3D bowtie shape is nothing but the surface formed by $B(\varphi, \gamma)$, where $\varphi$ is both the angle for the source rotation and the angle for the bowtie rotation around the axis perpendicularly through the center of the bowtie.

To calculate the contour of the bowtie, the coordinate systems are established in Figure 2, where $\sigma := \{o; x, y, z\}$ is fixed with the object, $\sigma_0 := \{o_0; x_0, y_0, z_0\}$ with the source, and $\sigma_b := \{o_b; x_b, y_b, z_b\}$ with the bowtie. Then, the contour of the bowtie can be expressed in $\sigma_b$ as

$$\begin{bmatrix} x_b \\ y_b \\ z_b \end{bmatrix} = \begin{bmatrix} S_0 \times \tan\gamma + \frac{B(\varphi,\gamma)}{\mu_b} \times \sin\gamma \\ \frac{B(\varphi,\gamma)}{\mu_b} \times \cos\gamma \\ 0 \end{bmatrix}, \tag{8}$$

where $S_0$ is the distance from the source to the bowtie, and $\mu_b$ is the attenuation coefficient of the bowtie. In the coordinate system $\sigma_0$, the contour of the bowtie is described as

$$\begin{bmatrix} x_0 \\ y_0 \\ z_0 \end{bmatrix} = \begin{bmatrix} \cos\varphi & 0 & -\sin\varphi \\ 0 & 1 & 0 \\ \sin\varphi & 0 & \cos\varphi \end{bmatrix} \begin{bmatrix} x_b \\ y_b \\ z_b \end{bmatrix} + \begin{bmatrix} 0 \\ S_0 \\ 0 \end{bmatrix}. \tag{9}$$

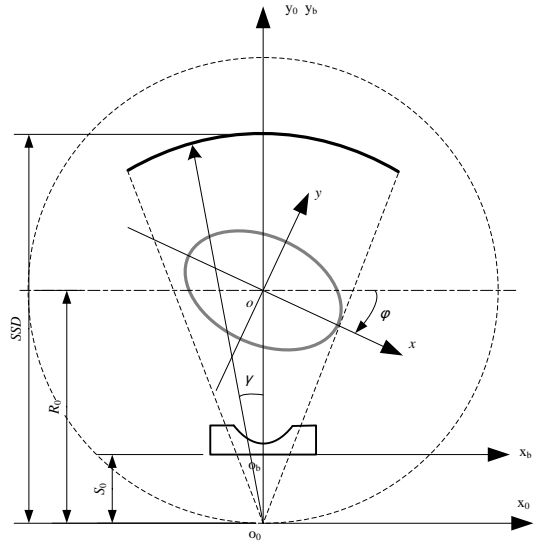

Figure 2: Coordinate systems. Both the object and the bowtie are rotated clockwise.

## 3. Numerical Results

### 3.1. Experimental Design

The parameters used for the design of an exemplary dynamic bowtie are listed in Table 1. They are representative of a commercial medical CT scanner [16]. The bowtie was made of aluminum, and intended for an ellipse water cylinder of semi-major axis $A$ = 200mm and semi-minor axis $B$ = 160mm.

Table 1: Parameters used for the design of a dynamic bowtie.

| Parameter | Value |
| --- | --- |
| Source trajectory | Full circle |
| Scan radius | 57cm |
| Source to detector distance($SSD$) | 10.4cm |
| Number of projections | 1160 |
| Number of detectors | 672 |
| Detector angular aperture ($\Delta\gamma$) | $1.354 \times 10^{-3}$ radians |
| FOV radius($R$) | 20cm |
| Water attenuation($\mu_w$)[22] | 0.183 cm$^{-1}$ |
| Bowtie attenuation($\mu_b$) [23] | 0.540 cm$^{-1}$ |
| Bowtie central thickness | 0.20cm |
| Semi-major axis of the ellipse ($A$) | 20cm |
| Semi-minor axis of the ellipse ($B$) | 0.8×A |
| Source to bowtie distance ($S_0$) | 15cm |

Figure 3 shows the projection profiles $P_w(\varphi, \gamma)$ of the elliptical water phantom without any bowtie filter. Setting the fan angle $\gamma = 0$ and the projection angle $\varphi = \pi/2$, the maximum projection value along the central ray $P_w(\pi/2, 0)$ can be obtained. Setting $\gamma = 0$ and $\varphi = 0$, the minimum projection value along the central ray $P_w(0, 0)$ can be obtained.

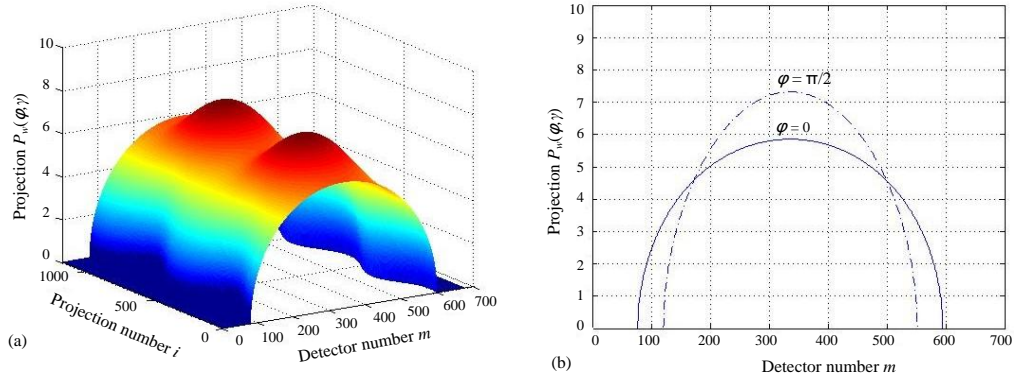

Figure 3: Projections of an elliptical water phantom. The projection angle $\varphi$ is indexed by the projection number $i$ ($0 \leq i < 1160$), $\varphi = i \times \Delta\varphi$, $\Delta\varphi = 2\pi/1160$. The ray angle $\gamma$ is indexed by the detector number, $\gamma = (335-m) \times \Delta\gamma$, ($0 \leq m < 672$). (a) A surface display for the sinogram of the water phantom, and (b) the extreme projection profiles of the water phantom for $\varphi = 0$ and $\varphi = \pi/2$ respectively.

Let $B(\varphi, 0) = 0.2 \times \mu_b$ in Eq. (5), which means that the bowtie central thickness is 0.2cm. With Eq. (7), the contour of the bowtie $B(\varphi, \gamma)$ is computed and visualized in Figure 4.

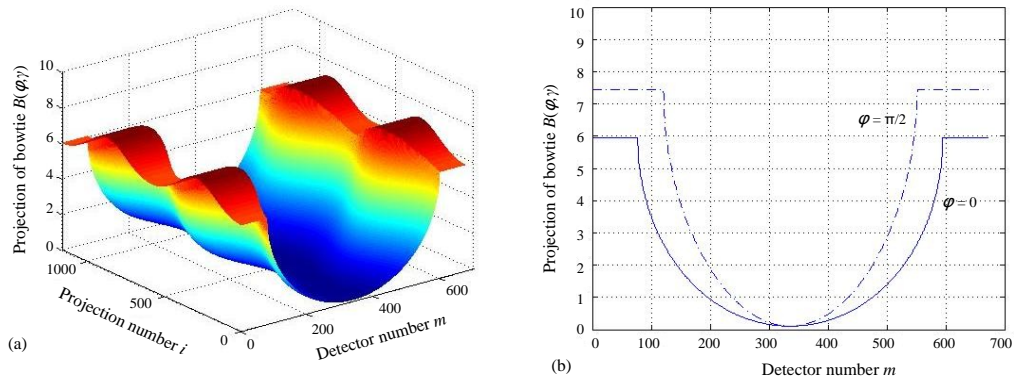

Figure 4: Visualization of the dynamic bowtie. (a) A surface display for the projections of the dynamic bowtie, (b) two projections of the bowtie for $\varphi = 0$ and $\varphi = \pi/2$ respectively.

As aforementioned, we view the dynamic bowtie as being shaped by rotating the angularly varying 2D contour around the axis perpendicularly through the bowtie center. The side facing the source is assumed flat for convenience. Then, based on the calculated $B(\varphi, \gamma)$ and

Eq. (8), we have the shape of the dynamic bowtie illustrated in Figure 5.

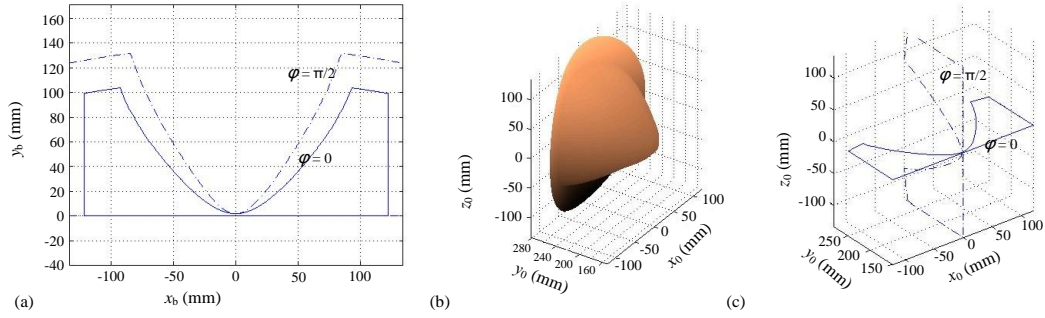

Figure 5: Shape of the dynamic bowtie. (a) The axial cross section profiles of the bowtie for $\varphi = 0$ and $\varphi = \pi/2$ respectively, (b) the curved surface of the bowtie, and (c) the 3D profiling of the bowtie through orthogonal cross sections at $\varphi = 0$ and $\varphi = \pi/2$ respectively.

### 3.2 Simulation in the Ideal Case

Next, let us examine how the dynamic bowtie impacts the radiation exposure throughout a field of view (FOV). First, the numbers of detected photons without the dynamic bowtie are calculated for the elliptical water phantom of $A = 200$mm and $B = 160$mm for a full scan, which are plotted in Figure 6. The number of emitted photons was set to $N_i = 200{,}000$ for each fan-beam ray.

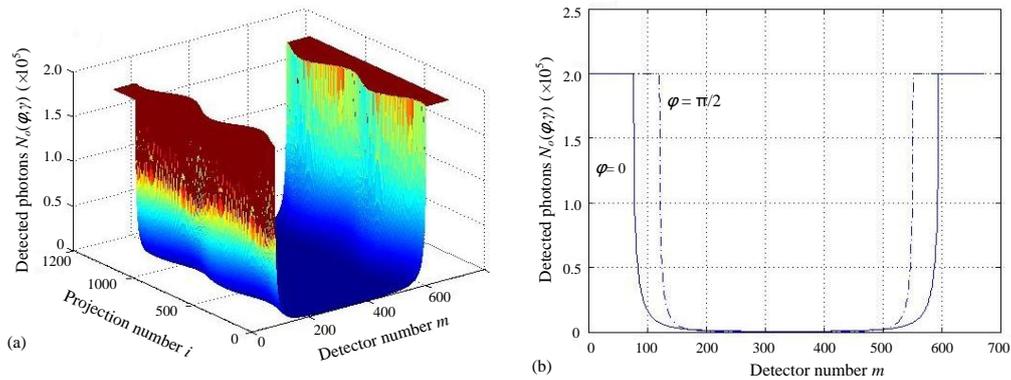

Figure 6: Visualization of numbers of detected photons through the elliptical water phantom without any bowtie. (a) A surface display of detected photons during the full scan, and (b) representative plots for $\varphi = 0$ and $\varphi = \pi/2$ respectively.

Second, with the dynamic bowtie for the aforementioned phantom and assuming $N_o = 100{,}000$, the number of photons can be simulated for each detector element. Figure 7 shows the profile of $N_i(\varphi)$ and the numbers of detected photons during the full scan. It can be seen that the numbers of detected photons are now quite uniform, which means the dynamic bowtie and the modulated current are nearly perfect for the elliptical water phantom.

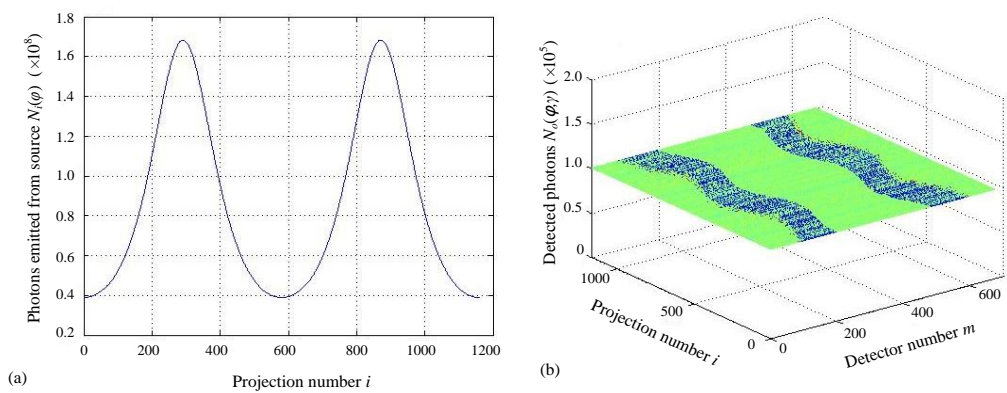

Figure 7: Visualization of numbers of detected photons through the elliptical water phantom with the dynamic bowtie. (a) A plot of the number of photons emitted from the source, and (b) the distribution of detected photons during the full scan.

Finally, we estimated the detected photons with a fixed bowtie under the same imaging conditions with the same water phantom for comparison. The fixed bowtie was designed for a circular water phantom of radius 200mm. The cross section of the bowtie and the resultant numbers of detected photons are shown in Figure 8. It is clear that the fixed bowtie is optimal for a circular phantom but it cannot equalize detected photons for a general object such as an ideal elliptical phantom.

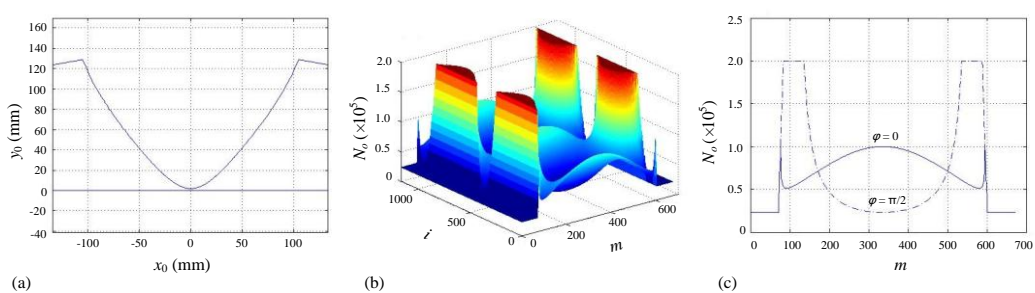

Figure 8: Visualization of numbers of detected photons through the elliptical water phantom with a fixed bowtie. The maximum number of detected photons was set to $2 \times 10^5$. (a) The cross section of the fixed bowtie, (b) the distribution of detected photons during the full scan, and (c) representative plots for $\varphi = 0$ and $\varphi = \pi/2$ respectively.

### 3.3. Simulation in a Practical Case

To show the merit of our dynamic bowtie, we randomly selected a CT slice and compared the dynamic ranges of the signals associated with the dynamic bowtie, a fixed bowtie, and no bowtie respectively. Figure 9 is a head CT image of 201mm width and 157mm height, which was approximated as an ellipse of semi-major axis $A = 100$mm and semi-minor axis $B = 80$mm.

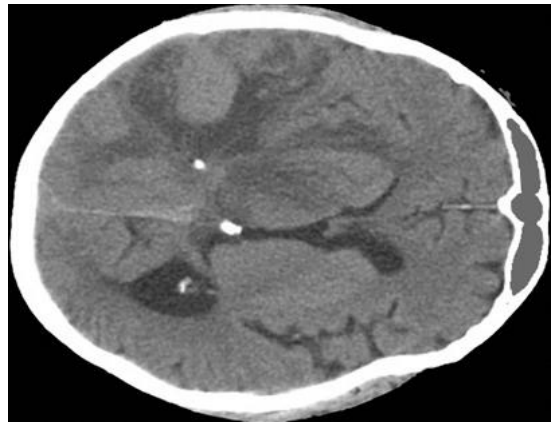

Figure 9: Head CT image approximated as an ellipse of semi-major axis A = 100mm and semi-minor axis B = 80mm.

As mentioned already, we designed a dynamic bowtie for the elliptical water phantom of $A = 100$mm and $B = 80$mm and a fixed bowtie for a circular water phantom of radius 100mm. Figure 10 illustrates the detected photons and representative profiles with and without the bowtie effect respectively. It can be seen that the dynamic ranges of the signals through the region of interest (ROI) differ greatly in the three cases, clearly showing the advantage of the dynamic bowtie, as shown in Figure 10 (e) and (f).

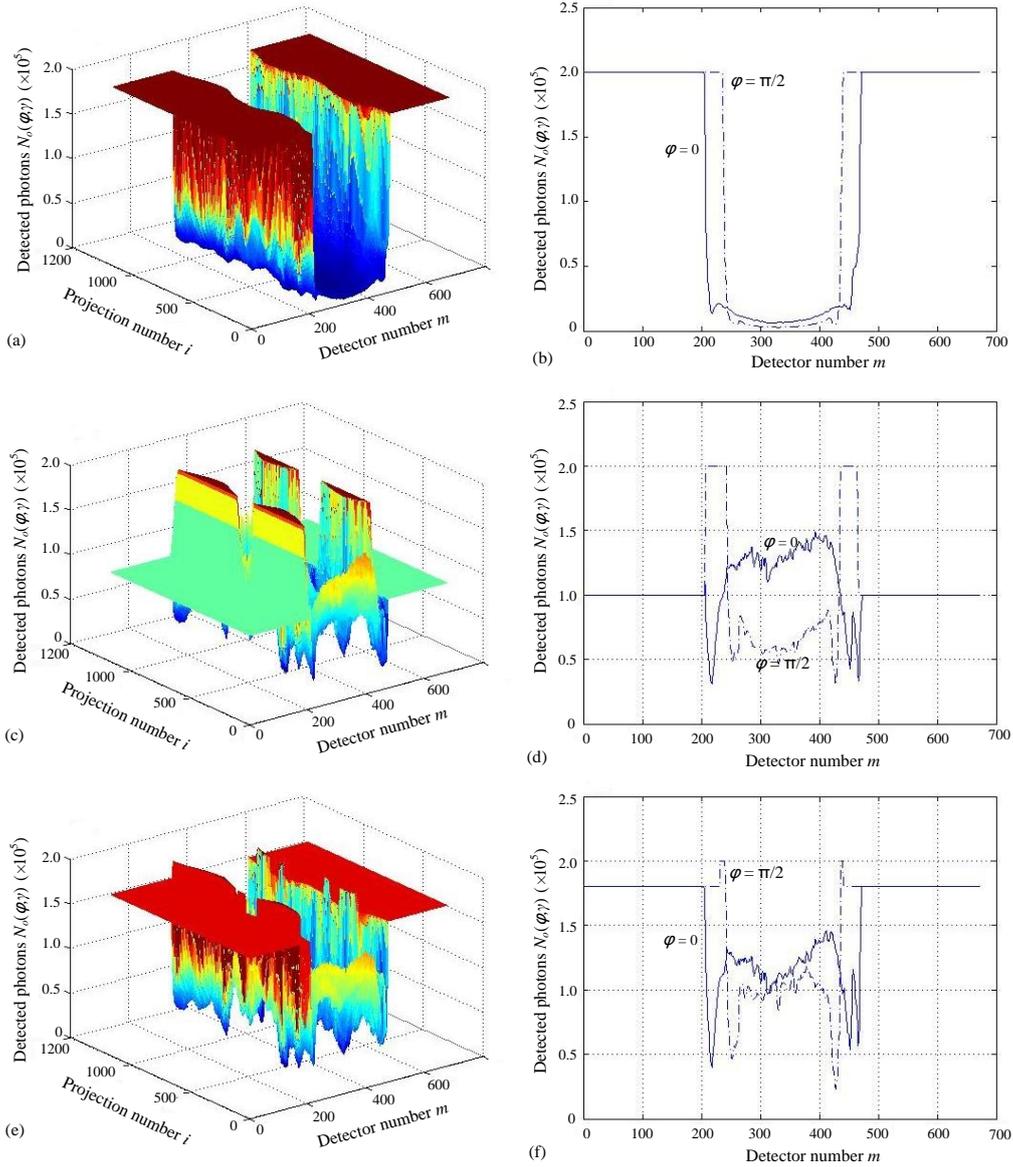

Figure 10: Detected photons and representative profiles for $\varphi = 0$ and $\varphi = \pi/2$ in the cases with and without the bowtie effect. The maximum number of detected photos was set to $2 \times 10^5$. (a) and (b) Without a bowtie, (c) and (d) with a fixed bowtie, (e) and (f) with the dynamic bowtie.

## 4. Discussions and Conclusion

As demonstrated in Figure 7, for an elliptical water phantom it is feasible to make the expected number of detected photons the same at each detector element by use of a dynamic bowtie and modulation of the x-ray tube current. In this way, if a cross section of a patient can be well approximated by an elliptical object, the dynamic range of detectors can be effectively matched to that of projection data. As a matter of fact, as long as a representative shape of cross sections is known, our methodology can be applied to design a dedicated dynamic bowtie to improve contrast resolution of the CT scanner.

As shown in Figure 10，without a bowtie, the information cannot be fully extracted from the FOV region since only a small portion of the detector dynamic range covers the signal variation. With use of a fixed bowtie, the detector dynamic range is better matched to that of projection data but data overflows are observed through the peripheral of the FOV, and the dynamic range of projection data is quite large. On the other hand, with a dynamic bowtie both of the data overflow and dynamic range problems are largely remedied.

As can be perceived in Figure 10 (e) and (f), the overflows of projection data through the peripheral of the FOV is due to poor registration of the assumed elliptical water phantom and the practical CT slice. Indeed, the difference between the assumed ellipse and the CT slice is primarily distributed along the peripheral of the patient support. This remaining data overflow challenge can be addressed by adding two dynamically sliding prisms on the two ends of the above-described bowtie, as shown in Figure 11. Further investigation along this line will be reported in a follow-up study.

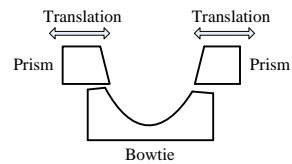

Figure 11: Sketch map illustration of the two prisms used to remedy the peripheral overflow.

It is acknowledged that in the current dynamic bowtie design，a monochromatic x-ray source has been assumed. In practice, an x-ray source is polychromatic. The multi-energy spectrum of x-rays introduces an additional layer of complexity. In this scenario, we need to augment the objective function in terms of numbers of detected photons and utilize the least square criterion for an overall optimization.

The proposed methodology for the design of our dynamic bowtie is to fit a rotating attenuator to a specific ellipse or an individual patient cross section. As shown above, a dynamic bowtie optimized for a water ellipse would not perform equally well on a water disk. This observation suggests that the rotating bowtie should be individualized. There are multiple ways to match a bowtie and a cross-section for fan-beam scanning. As an approximate option, Figure 11 has illustrated the two prisms to remedy the peripheral overflow. The associated adaptive translation of the sliding prisms helps x-ray flux equalization.   As the optimal solution, we make individualized dynamic bowties using rapid prototyping technology based on optical surface input. With future prototyping techniques in mind, it is not impossible that this process can be completely automatic and done in less than a couple of minutes.   We can simply capture the surface of a patient, deform a digital atlas such as the visible man dataset into the contour, estimate projections through a cross-section, and produce an ideal rotating bowtie. As far as helical scanning is concerned, a rotating bowtie can be designed in the same spirit. Specifically, with optical surface scanning and elastic deformation of a digital atlas we can estimate projections from an intended helical scan. Then, we can map the angular range of the helical scan onto a 360 degree interval for rotation of our dynamic bowtie. In other words, the dynamic bowtie rotation and the helical scan are fully synchronized from the beginning to the end. This scheme might be mechanically easier than the idea of printing a profile using cerium ink or metal component and sliding it across the source during the helical scan. The number of materials we can print is limited but there might be alternatives such as laser engraving. It is underlined that the adaptability of the dynamic bowtie promises significant performance improvement and dose saving, although bowtie planning is an additional overhead.

How do we generalize our dynamic bowtie work from fan-beam geometry to the cone-beam case? Our main idea is to replace the rotation of the bowtie with its translation. More specifically, given an appropriately shaped bowtie for a cone-beam of x-rays we can form various x-ray intensity distributions of different widths by changing the relative distance between the source and the bowtie, as illustrated in Figure 12. Coupled with two dynamically sliding prisms on the two ends of the cone-beam dynamic bowtie, an equalization effect similar to that of the fan-beam dynamic bowtie is anticipated. Since this is beyond the scope of the current paper, we will save the work for the future.

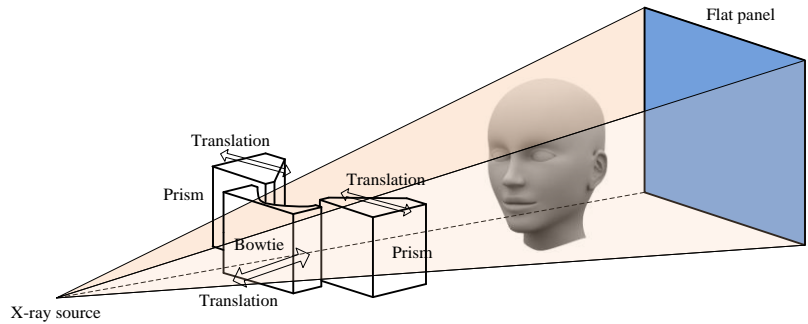

Fig 12: Illustration for the dynamically filtering of bowtie for cone-beam CT.

In conclusion, we have proposed a methodology for design of a dynamic bowtie, and demonstrated its feasibility in fan-beam geometry. Specifically, the bowtie is rotated in synchrony with the source rotation to optimize the x-ray intensity profile during the CT scan. Although the dynamic bowtie design has been done for an elliptical phantom, the resultant bowtie can be applied to a patient if his/her cross-section can be well approximated as an ellipse. Further refinement of the current design is under way. Also, this dynamic bowtie work will be extended to cone-beam geometry in the future.